\documentclass[10pt]{article}

\usepackage{pstricks}
\usepackage[margin=2.5cm]{geometry}
\usepackage{color}
\usepackage{graphicx}
\usepackage{amsmath} 
\usepackage{pifont} 
\usepackage{hyperref}
\usepackage[noadjust]{cite}
\usepackage{amssymb}

\usepackage[center,footnotesize,hang]{subfigure}

\newcommand{\sigmathermal}{\langle \sigma v\rangle}
\newcommand{\yes}{\ding{51}}
\newcommand{\no}{\ding{55}}
\newcommand{\yesno}{\ding{67}}

\begin{document}

\noindent \textbf{ \Large Theories relating baryon asymmetry and dark matter\\Mini review}\\

\noindent
S.M.Boucenna$^a$\footnote{ boucenna@ific.uv.es} and S.Morisi$^b$\footnote{ stefano.morisi@gmail.com}\\
$^a${ \it Instituto de F\'{\i}sica Corpuscular (CSIC-Universitat de Val\`{e}ncia), Apdo. 22085, E-46071 Valencia, Spain.}\\
$^b${ \it Institut f{\"u}r Theoretische Physik und Astrophysik, Universit{\"a}t W{\"u}rzburg, 97074 W{\"u}rzburg, Germany. }\\

\noindent \textbf{ Abstract}\\
\noindent \textit{  The nature of dark matter and the origin of the baryon asymmetry are two of the deepest mysteries of modern particle physics. In the absence of hints regarding a possible solution to these mysteries, many approaches have been developed to tackle them simultaneously { leading to very diverse and rich models}. We give a short review where we describe the general features of some of these models and an overview on the general problem. We also propose a diagrammatic notation to label the different models. }

\tableofcontents
\newpage

\section{Introduction}

The latest results on the cosmological parameters \cite{Ade:2013zuv} reveal that only 4.9\% of the content of the Universe is in the form of baryonic matter whereas 26.8\% is constituted by dark matter. The rest is accounted for by the mysterious dark energy. If we focus on the matter front then two disturbing questions are readily asked: \textit{ What is the nature of dark matter?} and \textit{ why is its density so close to the baryonic matter density, ie. $\Omega_{DM}\sim5 \, \Omega_{B}$?} 
%

Moreover, {  the above-mentioned visible matter density does not include anti-baryons ie. the visible universe is asymmetric with an initial excess of baryons over anti-baryons parametrized by  $\eta(b)=(n_b-n_{\overline{b}})/s\sim 10^{-10}$, where $n$ denotes the number density and $s$ the entropy density}. Therefore another fundamental question is \textit{ what is 
the origin of the observed baryon asymmetry of the universe (BAU)?}

This puts finding the nature of DM and the mechanism behind baryogenesis at the top of the agenda of modern physics \footnote{ For reviews on DM we refer the reader to \cite{Bertone:2004pz, Bergstrom:2000pn} and for baryogenesis to \cite{Dine:2003ax,Riotto:1999yt}.}. While the solutions to these two problems might well be unrelated to each other, it is nevertheless tempting to assume the new physics to be minimal and unifying enough so that it solves both of them with the same ingredients. Moreover if we discard simple numerical coincidence as an explanation to the intriguing closeness of matter densities, we are left with the task to construct theories relating them or unifying their genesis.

Indeed, numerous models have been proposed in the recent years to achieve this end. Broadly speaking, there are three approaches that are followed to relate dark matter to baryons. The first idea is that there is a sector connecting DM and baryons in the early universe. The connecting sector acts either as a parent sector, generating DM and baryons through decay for instance, or as a mediator mechanism transferring the asymmetry from the dark to the baryonic sector or vice versa. Asymmetric DM models (see below) used this approach extensively. The second approach uses the DM sector as an {  auxiliary} to a successful baryogenesis scenario. The strength of the phase transition in electroweak baryogenesis may for instance be enhanced by the presence of DM. The third approach uses the thermal WIMP paradigm as a framework to relate the abundances.

The purpose of this mini-review is to provide a succinct yet global picture on these models focusing on the key concepts and ingredients that are used in each reviewed model and on the predictions that are made.
While there are some similarities between these models, it is difficult to classify them in a consistent and easy way. Instead we opt for a diagrammatic approach Fig.\ref{fig:diag} and we review models that follow the main \textit{ roads} of the schematic. It is not our goal to be exhaustive with the references and we will refer to more systematic reviews when possible.

From the baryogenesis side we know that any mechanism that satisfies the three Sakharov condition\cite{Sakharov:1967dj}: B violation, C and CP violation and departure from thermal equilibrium can lead to a successful BAU. Whereas from the cold dark matter side we can generally speak of three classes of candidates:
weakly interacting massive particles (WIMPs), asymmetric dark matter (ADM) and non-thermal dark matter (NTDM)\footnote{ Where we include any non-thermally produced DM that does not fall in the ADM case.}.

In principle, we can organize the paper in terms of either one of these categories, we chose however to focus on the DM nature.

The paper is organized as follows. In section \ref{sec:wimp} we review models relating DM to the baryon asymmetry while preserving the WIMP \textit{ miracle}. Section \ref{sec:adm} is devoted to ADM models, where we will review different mechanisms and highlight the key concepts that are needed to construct them. In section \ref{sec:ntdm} we quickly mention the possibility of non-thermal DM. 
Finally we summarize the different models  and the roads taken  in Table \ref{table}. To simplify the understanding of the different models, we will specify in the text (in bold face) whenever it is helpful and in the table the path that is followed in the schematic. We will use the following convention: A star $*$ denotes the stage in the diagram where a new asymmetry appears while a bar on the top means that the direction of the arrow is flipped. We will also use the letter $T$ to refer to a thermalization stage.

\section{WIMP Dark Matter Models}
\label{sec:wimp}

It has been noted that relic particles from a thermal bath provide in a \textit{ miraculous} way the correct relic density of DM. Indeed, the number density of dark particles in the primordial thermal bath is \textit{ frozen-out} when the expansion rate  drops below the rate of the dark matter interactions. The abundance of the relic particle scales then as: 
\begin{equation}
\Omega_{\mathrm CDM} h^2 \simeq\frac{ 0.3  \times 10^{-26}~\mathrm{cm^3 s^{-1}} }{  \sigmathermal_{\mathrm f.o.}}, 
\end{equation}
{where $\sigmathermal_{\mathrm f.o.}$ is the thermal average of the annihilation cross-section of DM times the relative velocity at the time of freeze-out}.
Which gives the observed abundance for weak interactions cross-sections. This coincidence between DM and the weak scale has been dubbed the WIMP miracle. In addition to easily providing the observed relic abundance of DM, the WIMP paradigm is falsifiable. It offers a very rich array of phenomenological tests from underground direct detection experiments to astrophysical signals passing by colliders.
Without any doubt, maintaining the success of the WIMP paradigm and extending it to related DM to the baryon asymmetry is an attractive possibility. In this section we review the main theories attaining this goal.

\subsection{Electroweak baryogenesis}
\label{ewpht}
Electroweak baryogenesis is an appealing minimal scenario of baryogenesis based on the realization of the third Sakharov condition at the electroweak phase transition, see \cite{Riotto:1999yt} for a review on the mechanism. In the SM a strong first order phase transition, which is necessary in this scenario, requires a very light higgs boson ($<42$GeV), moreover the amount of CP violation in the SM is not enough to accommodate the observed BAU. These two considerations imply the need for new physics in order to have a successful baryogenesis and this is where DM comes in. The idea is to use the DM itself (or the dark sector particles) to make this scenario compatible with the SM higgs. A minimal extension of the SM with an extra (complex) scalar \cite{Profumo:2007wc,Espinosa:2011ax,Barger:2008jx,Cline:2012hg,Ahriche:2012ei} or two charged singlets \cite{Ahriche:2013zwa} achieves this goal, although recent data from LHC and WIMP direct detection experiments render this possibility less attractive because such a DM would have to be sub-dominant (ie. cannot account for the total density of DM). The same applies for inert higgs extensions of the SM \cite{Cline:2013bln} (higher $SU(2)$ representations were considered in \cite{Chowdhury:2011ga,AbdusSalam:2013eya}). However models with vector-like fermions are able to produce the total DM density and BAU for a wide range of masses \cite{Fairbairn:2013xaa}.

An even more extended higgs sector, say a 2-higgs-doublet model improves further the prospects of this scenario by providing the needed CP phases \cite{McLerran:1990zh,Cline:2011mm}. There is no direct correlation between DM and baryonic abundances in such theories, however the presence of the dark sector is necessary to have a successful baryogenesis which at the same time constrains the DM mass and couplings. {  Lastly, LHC and WIMP direct detection experiments may be used to constrain or rule out such a possibility. We note in passing that there are also models based on leptogenesis that follow the same philosophy outlined here, as in \cite{Basso:2012ti, Canetti:2012vf, Canetti:2012kh}.}

 \subsection{WIMPy  baryogenesis}
\label{wimpy}
Another possibility linking WIMP DM to the baryon asymmetry is the WIMPy baryogenesis model \cite{Cui:2011ab}. 
Here the baryon asymmetry arises from WIMP annihilation  instead of the decay of some heavy state like  for instance in the usual leptogenesis
mechanism.
It has been remarked that the annihilation of DM in the early universe can satisfy the Sakharov conditions and leads to a net baryon asymmetry
and the observed WIMP relic density. 

The baryon asymmetry generated with the WIMP annihilation can be washout from two kind of processes: inverse annihilation of baryons into DM
and baryon to antibaryon processes. Therefore the main requirement for any available WIMPy baryogenesis scenario is that washout processes must 
freeze-out before that WIMP freeze-out. Inverse annihilations are Boltzmann suppressed for $T< m_{\mathrm DM}$ but baryon to antibaryon washout
can be relevant also for $T\ll m_{\mathrm DM}$. One way to suppress such a processes is by introducing an exotic heavy antibaryon $\psi$ to which WIMP annihilate through the process $DM DM\to B\,\psi$ where $B$ is a SM baryon. If the exotic antibaryon $\psi$ has mass $m_\psi> m_{\mathrm DM}$, for $T< m_{DM}$
its abundance is Boltzmann suppressed and therefore the   baryon to antibaryon washout processes are suppressed. So the condition is 
\begin{equation}
m_{\mathrm DM}\lesssim m_\psi\lesssim 2 m_{\mathrm DM} 
\end{equation}
where the last condition comes from kinematic.
B (L) violation is achieved by annihilating the DM to two sectors: baryons (leptons) and exotic antibaryons (antileptons) that are individually asymmetric but together symmetric. It is important that the decay of the exotic particles do not erase the baryon asymmetry generated in the SM sector. For this
extra symmetry is required to decouple the exotic fields from the SM.

Solving the model-independent Boltzmann equations 
for the WIMPy baryogenesis framework, it is possible to show that the baryon asymmetry is proportional to the DM density at the time 
of freeze-out of the washout processes, i.e.\footnote{Here  $Y_X$ is the ratio of the number density $n_X$ of the specie 
$X$ with the entropy $s$.}
\begin{equation}\label{YB}
Y_{\mathrm B}\approx \frac{\epsilon}{2} [ Y_{{DM}_{\mathrm wo}}-Y_{DM} ]
\end{equation}
where $Y_{{DM}_{\mathrm wo}}$ is the DM density at the washout while $Y_{\mathrm B,{DM}}$ are the observed baryon and DM densities and $\epsilon$ 
is the baryon-antibaryon asymmetry.
%
From eq.\,(\ref{YB}) and the relation 
\begin{equation}
Y_{DM}= \frac{(5\,GeV)}{m_{\mathrm DM}}\, Y_{\mathrm B} 
\end{equation}
it follows that  $Y_{{DM}_{\mathrm wo}}\gg Y_{DM}$, namely 
it is crucial to freeze-out the wash out processes \textit{ before} the WIMP freeze-out temperature otherwise any generated asymmetry would be quickly erased.

\vskip5.mm

As a concrete example we consider a realization of the WIMPy idea in which the WIMP annihilate to leptons
generating a lepton asymmetry then converted into a baryon asymmetry through sphaleron like in leptogenesis. 
The DM candidate consists of a pair of gauge singlet Dirac fermions $Y$ and $\bar{Y}$. In addition to DM two new weak-scale states $\psi$ (fermion $SU_L(2)$ doublet) and $S_1$ and $S_2$ (pseudo-scalar gauge singlets) are added. 
The fields $\{ Y,\overline{Y}, \psi,\overline{\psi},S_i  \}$ transform under an extra $Z_4$ symmetry respectively as $\{ +i,-i, -1,-1,-1 \}$.
The Lagrangian contains the extra terms
\begin{equation}
\mathcal{L}\supset (\lambda_i Y^2+\lambda'_i\overline{Y}^2)S_i + \lambda_{\psi_i} L\psi S_i\,. 
\end{equation}
Since there is more than one scalar $S_i$, it remains a relative complex phase between the $\lambda$ couplings.
Then as in the common leptogenesis case the interference between tree level and loop diagrams give rise to CP violation resulting in 
an asymmetries in $L$  (\textbf{ 4$^*$})\footnote{ The depiction of this step in the schematic:  
the DM annihilates into the visible sector (line 4), with the * is there to show that an asymmetry is produced in this step.}
and subsequently converted to B asymmetry by means of the sphalerons.
Here differently from leptogenesis,  the dark matter $Y$ annihilates into SM leptons $L$ and  $\psi$ through the pseudo-scalars (\textbf{ T})\footnote{We use here the letter T to emphasize that the DM is thermally produced.} then a lepton asymmetry also accumulate in $\psi$.
The processes linking $\psi$ to the SM do not erase the lepton asymmetry thanks to the extra $Z_4$ symmetry that decouples $\psi$ from the SM. 

At the end an asymmetry is generated from a $2\to 2$ process instead of a $1\to 2$. An important requirement is that $m_\psi > m_Y$ 
because it implies that the dominant washout process $L\psi \to L^\dagger \psi^\dagger$ is Boltzmann-suppressed when DM
is annihilating. We summarize diagrammatically the  signature of the model as (\textbf{ T--4$^*$}), as it appears in the Table \ref{table}.

The detection prospects are rich in this scenario and include direct (for models with annihilation to quarks), indirect detection (anti-deuteron) and collider signals. See \cite{Bernal:2012gv,Bernal:2013bga} for a general phenomenological study of this class of models. Other models preserving the WIMP miracle and attempting to relate the DM to BAU can be found in Ref.\,\cite{Davidson:2012fn, McDonald:2011zza}.

\subsection{Meta-stable WIMP}
As in the case of WIMPy baryogenesis, this model \cite{Cui:2012jh} attempts to explain the DM/baryon relic density coincidence using the WIMP miracle. The idea is to use a decaying WIMP instead of a stable one.
A thermal WIMP $Y$ freeze out at a temperature $T_f$ that is typically $T_f\sim m_Y/20$. At freeze out the WIMP density is $Y_Y(T_f)$
which is equal to the DM density today $Y_Y(T_f)\simeq Y_Y(T_0)$ if the WIMP is stable. The authors consider two kinds of WIMPs: one stable $Y_1$ that is the DM candidate and one $Y_2$ that decay after freeze out, with the densities of the two WIMPS at the freeze out being almost the same $Y_{Y_1}(T_f) \approx Y_{Y_2}(T_f)$. The density of the decay WIMP at the freeze out temperature is the initial condition for the baryogenesis. 

The meta-stable WIMP $Y_2$  decays after thermal freeze-out into baryons in such a way that the baryon number B and CP are violated. In a minimal realization of the idea, the SM is extended to include a di-quark scalar $\phi$  and $\psi$ which are Majorana fermions and a singlet scalar $S$. The relevant couplings are 
\begin{equation}
\phi dd \,,\quad Y_2\bar{u}\phi\,,\quad \psi \bar{u}\phi\,,\quad Y_2^2 S\,,\quad |H|^2 S 
\end{equation}
Where $u$ and $d$ are the SM quarks. The scalar $S$ mediates the thermal annihilation of $Y_2 Y_2$ into SM.
The meta-stable WIMP decay as $Y_2\to u\phi^*$ followed by the decay of $\phi \to dd$. 
A CP asymmetry $\epsilon_{CP}$  in $Y_2\to u\phi^*$ and $\overline{Y}_2\to \bar{u}\phi$  arises  from the interference between the tree-level diagram with the one loop diagram mediated by $\psi$ (that shares with $Y_2$  the same quantum numbers).
In order to generate a baryon asymmetry the WIMP must decay before the BBN and after WIMP freeze out, i.e.  $T_{BBN}< T_{Y_2}< T_f$.
Solving the Boltzmann equations it is possible to find the baryon density today
\begin{equation}
Y_B(T_0)\approx  \epsilon_{CP} \int_{T_0}^{T_D} \frac{dY_{Y_2}}{dT} dT \simeq  \epsilon_{CP} Y_{Y_2}(T_f)
\end{equation}
Using the relations  $Y_Y(T_f) \approx Y_{Y_2}(T_f)$ and that $Y_Y(T_f)\simeq Y_Y(T_0)$ we arrive at the result
\begin{equation}
\Omega_B=\epsilon_{CP}\frac{m_p}{m_{DM}}\, \Omega_{DM}
\end{equation}
where $\Omega_{DM}$ is the relic abundance of the DM. The model lies at the electroweak scale and therefore it can be probed in colliders.

\begin{figure}[t]
\centering
{\includegraphics[width=.85\textwidth]{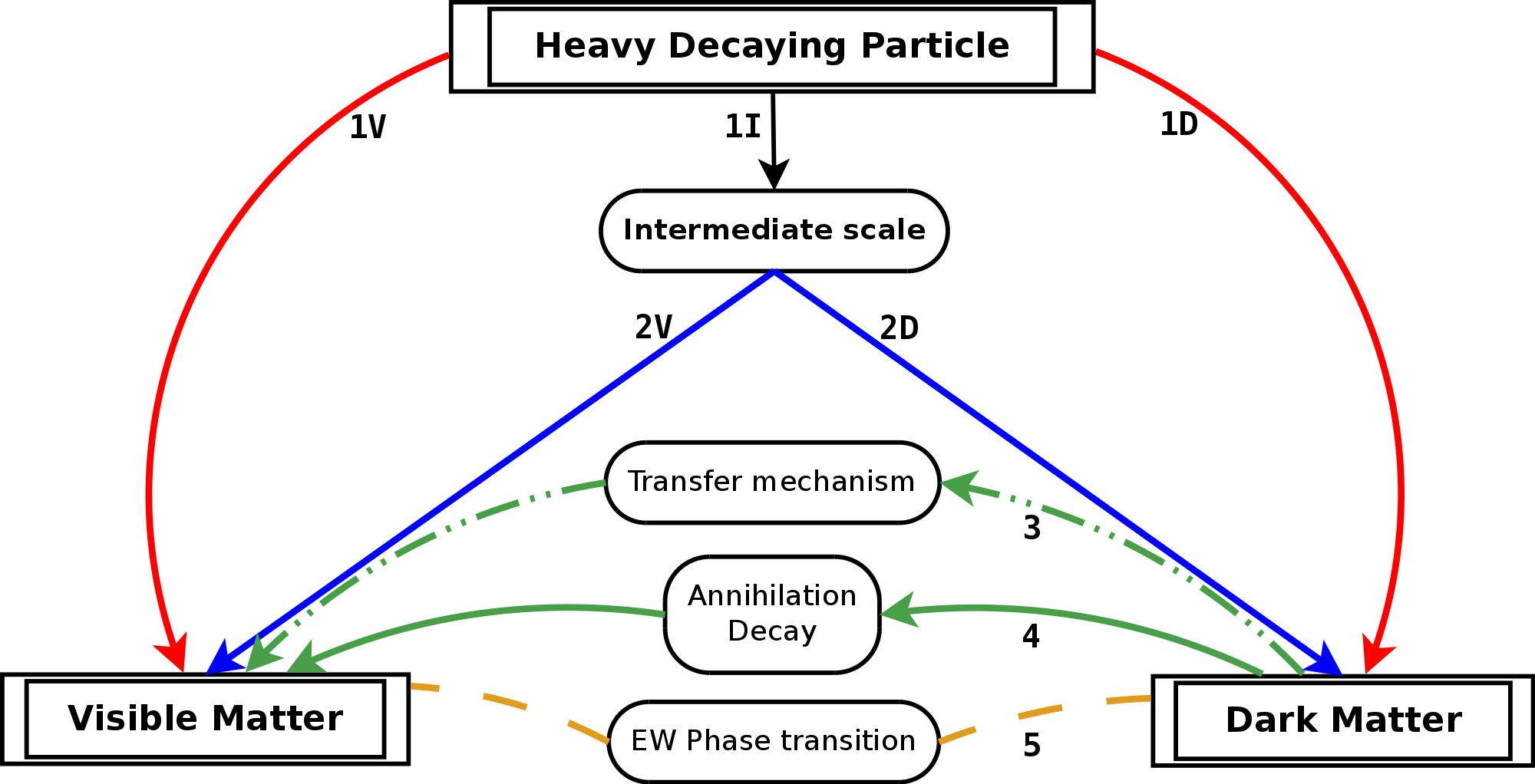} }
\caption{A schematic of the different mechanisms relating DM to baryon asymmetry. The lines are the different stages of the considered mechanism. The labels on the lines are used to describe the model. 
}
    \label{fig:diag} 
\end{figure}

\section{Asymmetric Dark Matter Models}
\label{sec:adm}

ADM \cite{Nussinov:1985xr,Barr:1990ca,Barr:1991qn,Kuzmin:1996he,Hooper:2004dc,Farrar:2004qy,Kitano:2004sv,Kaplan:2009ag} is a class of DM models often seen as an alternative to the WIMP paradigm. The rationale of ADM is based on the hypothesis that DM abundance is, similarly to baryons, only the surviving asymmetric part of the initial density and is of the same order as the baryon asymmetry, i.e. 
\begin{equation}
n_{Y}-n_{\overline{Y}}\sim n_b-n_{\overline{b}}
\end{equation}
where $Y$ denotes the DM particle. The motivation comes from the fact that the observed DM and visible matter abundances are remarkably close to each other. These models usually lead to a relation between DM mass and proton mass: $M_{DM} \sim 5 M_P$ in contrast with WIMP DM models where the scale of reference is the weak scale. The relation between the DM mass and the proton mass is however not explained except in some models based on hidden sectors such as in mirror worlds \cite{Foot:2004pq,Hodges:1993yb,Mohapatra:2000rk}, models with a dark QCD \cite{Bai:2013xga} or composite models (see below).

ADM can be implemented in many ways leading to a very rich theoretical and phenomenological landscape. While it is difficult to classify these models in a straightforward way, it is nevertheless enriching to highlight the key principles they usually rely on. Basically two main approaches are followed: 1) Dark and visible matter asymmetries are generated at the same time. This is usually achieved with the decay of a heavy particle. 2) The asymmetry is generated in the dark sector then is transferred (\textit{ via} sphaleron processes, higher dimension operators or renormalizable interactions) to the visible sector or vice versa.
It is also necessary to pass at some point by a thermalization phase to get rid or to avoid the production of the symmetric part of DM {  (a less extreme cancellation of the asymmetric part leads to mixed scenarios between WIMP and ADM \cite{Graesser:2011wi}).}

We will present here ADM models explicitly showing the key assumptions and principles used as well as their phenomenological impact. They make use of the main ADM concepts and pass by the main diagrammatic roads. For a recent review and an exhaustive list of reference we refer the reader to \cite{Sannino:2009za,Petraki:2013wwa,Zurek:2013wia} and for a more succinct overview \cite{Davoudiasl:2012uw}.

\subsection{Composite ADM}
\label{cadm}

The idea of the ADM has been proposed in the seminal work of Nussinov \cite{Nussinov:1985xr} who suggested that in analogy with the visible sector's baryon asymmetry, a technibaryon asymmetry is a natural possibility. This idea has been recently revamped in the context of walking dynamics \cite{Gudnason:2006ug,Gudnason:2006yj,Nardi:2008ix,DelNobile:2011je}. If the model is arranged such that the lightest technibaryon (LTB) is neutral and stable, 
the density of the LTB scales as:
\begin{equation}
\frac{\Omega_{TB}}{\Omega_{B}}=\frac{{TB}}{{B}}\frac{m_{TB}}{m_{p}}
\end{equation}  

Where $m_p$ is the proton mass, $m_{TB}$ is the mass of the LTB. TB and B are the technibaryon and baryon number densities, respectively.
This is the typical scaling of ADM models.

The model discussed in \cite{Gudnason:2006ug} is a technicolor theory based on the $SU(4)$ global symmetry spontaneously broken down to $SO(4)$. 
Such a breaking gives rise to 9 Goldston bosons, three of them corresponding to the SM gauge bosons. The remmant six Goldstone bosons carry technibaryon charge and the lightest of them (LTB) is the DM candidate\footnote{ The Goldstone bosons are supposed to pick up a mass from a higher scale.}.
%
%
In \cite{Lewis:2011zb,Hietanen:2013fya} the properties of composite (asymmetric or symmetric) dark matter candidates have been computed in detail via first-principle lattice simulations.

%

\subsection{Kitano-Low}
\label{kitanolow}

The model implemented in \cite{Kitano:2004sv} considers a mechanism originally proposed in \cite{Kuzmin:1996he} to unify in an elegant way the abundances of DM and baryons. It is a prototype of the ADM models based on decay of a field connecting the dark and visible sectors.

The authors postulate a new symmetry, namely a $Z_2$ parity, under which the SM particles are neutral and new particles are charged,
forming a dark or hidden sector. The lightest of the hidden particle is stable and is a DM candidate. 
A generalized B-L number is unbroken and is shared between the SM and the dark sector, 
thus any excess of B-L that is generated in one of the two sectors is compensated by the same excess in the other sector. After baryogenesis the interactions between the visible and the dark sectors become negligible and the B-L excesses are separately conserved in the two sectors giving a relation between the visible and dark relic densities.

A simple model realizing the idea consists of a heavy particle $P$, a messenger particle $X$ which carries a color charge and the DM candidate $Y$, all odd under the $Z_2$ while the SM is even. 
The mechanism passes through 3 stages. In the first stage $P$  has CP-violating out of equilibrium decays into
 SM  and to a lighter messenger $X$ generating an excess in both sectors but preserving the generalized B-L globally.
Then is assumed that below the baryogenesis temperature the two sector are decoupled and the two asymmetries are conserved such that we have:
\begin{equation}
n_{\mathrm B-L}^{\mathrm SM} = -n_{\mathrm B-L}^{ X} \sim n_{X}-n_{\bar{X}}
\end{equation}
In the second stage the dark $X$ messenger annihilate away its symmetric part with $\overline{X}$ through gauge interactions 
and we are left with its asymmetric part only.
In the third and final stage the decay of $X$ to DM particle $Y$ and therefore 
\begin{equation}
n_{\mathrm DM} \propto n_{\mathrm B-L}^{ X}
\end{equation}
giving a tight relation between the visible (baryonic) and DM number densities.
To ensure that such a relation exists it is important  that $X$ is long lived enough such that it decays after its symmetric part cancels out.

%
%
%

We summarize the mechanism: Decay of $P$ that produces the asymmetries in $X$ (\textbf{ 1$_I^*$}) and SM (\textbf{ 1$_V^*$}, since an asymmetry in the visible sector is generated by the decay) followed by symmetric annihilation of $X$ (\textbf{ T})
and finally the decay of $X$ to  the lightest dark particle  $Y$ (\textbf{ 2$_D$}). We denote the full mechanism in a compact way as
(\textbf{ 1$_V^*$--1$_I^*$--T--2$_D$})

An interesting possibility is to consider $X$ itself as the DM particle. This possibility is not possible here here because of charge assignment of $X$ (colored particle). However, we will see now that Hylogenesis realizes this possibility. Note that the original asymmetry can be generated through 
the Affleck-Dine \cite{Affleck:1984fy,Allahverdi:2012ju} mechanism in a SUSY framework \cite{Bell:2011tn,Cheung:2011if,vonHarling:2012yn,Roszkowski:2006kw} or through leptogenesis as in \cite{Cosme:2005sb}.

\subsection{Hylogenesis}
\label{hylogenesis}

This model \cite{Davoudiasl:2010am} is based on a hidden sector composed of 3 Dirac fermions $X_1$, $X_2$, $Y$ and a complex scalar $\phi$. It is assumed that 
$M_\phi \sim M_Y\sim GeV$ and $TeV<M_{X_1}<M_{X_2}$. 
$X_i$ are made to couple to the visible sector through the \textit{ neutron portal} ($X_i \,d^c u^c d^c$),  the relevant terms in the Lagrangian  are:  
\begin{equation}
\mathcal{L} \supset  \frac{\lambda_i}{\Lambda^2}\,X_i d^c u^c d^c+ \kappa_i \, \bar{X}_i Y \Phi + \textrm{h.c.}
\end{equation}
The particle content and the symmetries of the model permit the definition of a generalized baryon number (B), conserved by both sectors, 
under which $B_X = -(B_Y+B_\Phi)=1$ as well as non-reducible CP phases.

In the early universe an equal number of $X_1$ and its antiparticle $\overline{X}_1$ are generated nonthermally (e.g. during reheating) 
and the total baryon number is zero at this stage. 
Then both states   $X_1$ and $\overline{X}_1$ decay into the visible and hidden states as 
%
$X_1 \to udd$ (\textbf{ 1$_V^*$}) and $X_1 \to \bar{Y} \Phi^*$  (\textbf{ 1$_D^*$}) and their conjugates
at tree level and through loops (including the lighter dark particles $\phi$ and $Y$), generating
an asymmetry in the visible sector $\epsilon_V$ and an asymmetry in the hidden sector $\epsilon_D$ \textit{ \'a la} leptogenesis
\begin{equation}
\epsilon_V = \frac{\Gamma(X_1 \to udd)-\Gamma(\overline{X}_1 \to \bar{u} \bar{d} \bar{d})}{\Gamma_{X_1}}\simeq 
\frac{m_{X_1}^5Im(\lambda_1^*\lambda_2\kappa_1\kappa_2^*)}{256 \pi^3 |\kappa_1|^2\Lambda^4 m_{X_2}}.
\end{equation}
where $\Gamma_{X_1}$ is the total rate and $\epsilon_D$ can be obtained in a similar way. We have
\begin{equation}\label{eq14}
\begin{array}{lcl}
\Gamma(X_1 \to udd)&=&\Gamma_V+\epsilon_V\Gamma_{X_1},\\
\Gamma(\overline{X}_1 \to \bar{u} \bar{d} \bar{d})&=&\Gamma_V-\epsilon_V\Gamma_{X_1},\\
\Gamma(X_1 \to \bar{Y} \Phi^*)&=&\Gamma_D-\epsilon_D\Gamma_{X_1},\\
\Gamma(\overline{X}_1 \to Y \Phi)&=&\Gamma_D+\epsilon_D\Gamma_{X_1}.
\end{array}
\end{equation}
Because $X_1$ is a Dirac particle, the asymmetry generated in the visible sector is then translated as an asymmetry in the hidden sector. Indeed, CPT invariance forces the particle and its anti-particle to have equal total decay rates $\Gamma[X_1\to n +\bar{Y} \Phi^*] = \Gamma[\bar{X_1}\to \bar{n} +Y \Phi] $, which translates as a relation between asymmetries, that is $\epsilon_D = -\epsilon_V$ where we have used the eq.\,(\ref{eq14}). Therefore in the decay of  $X_1$ and $\overline{X}_1$
a baryon number is generated in the visible sector and an equal and opposite baryon number is generated in the hidden sector so that the total
baryon number is zero.
The two asymmetries are frozen-in thanks to the weakness of the interactions between the two sectors.

The final step is to cancel out the symmetric part of the dark matter particles and this is achieved for instance with an extra $U(1)_D$ gauge symmetry in the hidden sector under which $Y$ and $\Phi$ have opposite charges and $X_{1,2}$ are neutral. The symmetric part is depleted (\textbf{ T}) by the annihilation
processes $Y\overline{Y} \to Z' Z'$ and $\Phi {\Phi}^* \to Z' Z'$ with $m_{Z'} < m_{Y,\Phi}\sim GeV$ (this is consistent with present observations for
$10^{-6}<\kappa<10^{-2}$) with $Z'$ decaying to SM through photon. These cross section are much larger to the one need to obtain the correct
DM relic density by thermal freeze-out. Then the DM density is given by the residual asymmetric component 
and we are then left with the relation:
\begin{equation}
n_Y =n_\Phi=n_B
\end{equation}
that gives a strong relation between the visible and dark matter abundances:
\begin{equation}
\frac{\Omega_{DM}}{\Omega_B} = \frac{(M_Y+M_{\Phi})}{M_P }\sim 5\,.
\end{equation}
We denote in a compact way this mechanism with the signature (\textbf{ 1$_V^*$--1$_D^*$--T}). 
Because of the neutron portal, hylogenesis provides an interesting signature of the DM: the induced proton decay (IND). Indeed DM can scatter with protons producing mesons $\phi^*p \to Y K^+ $.

\subsection{ADM from leptogenesis}
\label{admlepto}
If we take Majorana instead of Dirac decaying fields in the previous model, we get different consequences on the DM mass. The model considered in \cite{Falkowski:2011xh} is based on the decay of a heavy right handed neutrino field $N$. 

The model is an extension of the SM and consists of two right-handed neutrinos and a scalar $\phi$ and fermion $Y$ gauge singlets,
charged under an extra $Z_2$ parity, that made the hidden sector
\begin{equation}
\mathcal{L} \supset  M_i N_i^2 + y_i LH N_i + \lambda_i N_i Y\phi+ \textrm{h.c.}
\end{equation}
$N$ couples to the SM with Dirac Yukawa coupling and to the hidden sector, 
$Y$ is the DM candidate. 
Therefore $N$ can decay (out of equilibrium) simultaneously as $N\to L H$ (\textbf{ 1$_V^*$}) and $N\to Y \phi$ (\textbf{ 1$_D^*$}) generating two different and unrelated CP-asymmetries $\epsilon_L$ and $\epsilon_{DM}$ respectively. Here $N$ is a Majorana particle and CPT does not imply that  $|\epsilon_L |= |\epsilon_{DM}|$ like in Hylogenesis (see previous section).
The DM must be a Dirac particle in order to preserve a lepton number. 
Because both $Y$ and $\phi$ are charged under the extra $Z_2$, the DM is stable and the hidden sector can interact with the SM only by means of the heavy right-handed neutrino. In order to cancel out the symmetric component of the DM, an additional gauged $U(1)$ interaction is imposed to annihilate the $Y, \,\overline{Y}$ pair. We are left with the asymmetric parts of $Y$ (\textbf{ T}). The DM and baryon density $\Omega_{DM}/\Omega_{B}$ is then proportional to the ratio of the  CP-asymmetries $\epsilon_{DM}/\epsilon_{L}$, 

\begin{equation}
\frac{\Omega_{\mathrm DM}}{\Omega_{B}} \sim \frac{m_{\mathrm DM}}{m_p} \frac{\epsilon_{DM}}{\epsilon_{L}} \frac{\eta_{DM}}{\eta_{L}}
\end{equation}
where $\eta_{DM,L}$ are the washout factor. Therefore the DM mass can be very different from the value of $5 M_p$ given in most ADM models. A similar model based on type-II leptogenesis instead of type-I has been proposed in \cite{Arina:2011cu}. See also \cite{Cosme:2005sb} for an earlier ADM model based on leptogenesis and where the DM mass is in the typical few GeV scale.

\subsection{Darkogenesis}
\label{darkogenesis}
In this model \cite{Shelton:2010ta} an asymmetry is generated in the dark sector and is then transferred to the visible sector. The DM asymmetry arises from a first order dark phase transition in the hidden sector to  which the SM does not participate. 
The dark baryogenesis proceed through the symmetry breaking phase transition of a dark non-Abelian gauge group $G_D$.
The fields in the dark sector have a global dark symmetry $U_D(1)$ which is anomalous under $G_D$. During the 
symmetry breaking first order phase transition an dark asymmetry is generated by means of CP violating interactions.


The asymmetry can be transferred to the visible sectors in two ways: by fields that carry both hidden and visible charges (perturbatively) 
or via electroweak sphalerons (nonperturbatively).
In the last case, in order to transmit the asymmetry from the dark sector to the SM one, it is required a mediator charged under 
both the $SU_L(2)$ and the dark symmetry $U_D(1)$. Then the dark number is anomalous under $SU_L(2)$ and the SM electroweak sphaleron
can convert the asymmetry of the dark sector into an asymmetry in the SM.
 
In the first case the connectors can consists of higher order effective operators of the type 
\begin{equation}
O_d\, LH,\quad O_d\, udd,\quad O_d\, LLe,\quad O_d \, LQd,\quad O_d\, LH LH, 
\end{equation}
where $O_d$ is a dark sector operator like for instance $O_d=X,\,X^2$. 
The hidden sector phase transition occurs at a temperature above the temperature at which the effective transfer operator freeze-out
The dark matter mass lies around 5 to 15 times the mass of the proton.

Direct detection cannot falsify the darkogenesis mechanism, however the gravitational wave signal from dark first order transition could in principle probe this mechanism. {  The asymmetry in the dark sector can also be generated via a different baryogenesis mechanism, see \cite{Feng:2013wn} for an example where a heavy particle decays to the dark sector, creating an asymmetry there that is then transferred to the visible sector. For the opposite case, see \cite{Feng:2012jn} or \textit{ aidnogenesis}\cite{Blennow:2010qp}, for instance where the asymmetry is transferred through sphalerons from the SM to the dark sector. Diagrammatically we denote this model as : \textbf{ *--3--*}, which means that an original asymmetry in the dark sector (following the direction the arrow) is transferred to the visible sector. See also \cite{Barr:2013tea} for a recent model where sphalerons are responsible for cogenerating the dark matter.
}

\subsection{Xogenesis}
\label{xogenesis}
Like in the \textit{ darkogenesis} model, here \cite{Buckley:2010ui} a DM asymmetry is created and then transferred to the baryon by means
of transfer operators. The problem of the creation of a DM asymmetry is not addressed here and the authors focuses on the transfer mechanisms. The main difference between this mechanism and the classic ADM ones going in the same direction is that the DM mass can be around the weak scale instead of the proton mass (for a different idea how to obtain heavy ADM see \cite{Gu:2010ft}) without fine-tuning the parameters. The main idea can be summarized as follows: {  If DM is not relativistic at the temperature where the transfer operator decouples $T_D$ then the DM number density undergoes a thermal suppression allowing the DM to be heavy.}

The transfer can be due from the $SU_L(2)$ sphalerons (or the exotic sphalerons of a new gauge group) or lepton/baryon number violation from higher order operators.  In any transfer scenario chemical equilibrium between DM and baryon is maintained until the transfer operator decouples. 
When the transfer is active, we have:
\begin{equation}\label{mumu}
\mu_{DM} \sim \mu_B\,.
\end{equation}
Given a specie $i$ in general its asymmetry  $n_{\Delta i}= n_i-\bar{n}_i$ is proportional to its chemical potential 
\begin{equation}\label{cmu}
n_i= c_i \mu_i
\end{equation} 
The coefficients $c_i$ are function of the mass and temperature $c_i = c_i  (m_i, T)$  \cite{Barr:1990ca}:
\begin{equation}
c_i = g_i f(m_i/T) T^2 R^3\,,\qquad f(x)= \frac{1}{4\pi^2}\int_0^\infty \frac{y^2 dy}{\cosh^2(\sqrt{x^2+y^2}/2)}
\end{equation}
where $g_i$ is the statistical weight and $R$ is the Robertson-Walker scale factor at temperature $T$. For small value of $m_i/T$
then $f(m_i/T)$ tend to a constant, while for large $m_i/T$ then $f(m_i/T)$ is very small 
\begin{equation}\label{fmu}
 f(m_i/T)\sim \left[
\begin{array}{l}
m_i/T\ll 1\,,\quad  1/6\\
m_i/T\gg 1\,,\quad 2(m/2\pi T)^{3/2}e^{-m/T}
\end{array}
\right]
\end{equation}
Typically only the first possibility where $m_{DM}/T_D\ll 1$ is taken ($T_D$ is the decoupling temperature of the transfer operator). 
In this case from eq.\,(\ref{mumu}) and\,(\ref{cmu}) 
it follows that $n_{\Delta_{DM}}\sim n_{\Delta_B}$ leading to the 'prediction' $m_{DM}\sim 5 \, m_p$. 
However, a second solution is possible. If the ratio $m_{DM}/T_D$ is large, then the coefficients
$c_{DM}$ is suppressed, see eq.\,(\ref{cmu}) and\,(\ref{fmu}). This results in a lower $n_{DM}$ with respect to the case where the ratio $m_i/T_D$ is small and thus a larger DM is allowed. For a given value of $T_D$ the non-relativistic solution give about $m_{DM}\sim 10\,T_D$
instead of 5GeV (relativistic solution), giving a mass for the DM of the order of the TeV.

A simple example is given by a DM particle $Y$ that transform as a fermion doublet of $SU_L(2)$ with hypercharge $+1/2$.
Since the DM is charged under $SU_L(2)$, it interacts with the SM sphaleron. Thanks to the sphaleron $Y$ and quarks are
in thermal equilibrium, therefore the DM and quarks chemical potential are releted, i.e. $\mu_Y=-3 \mu_{u_L}$. In this example
the decoupling temperature  $T_D$ of the transfer operator is the temperature where the spahleron is no more active, that is 
around 200\,GeV. Solving equations\,(\ref{mumu}) and \,(\ref{cmu}) one gets for the DM a value of about 2000\,GeV.

The idea has been illustrated with different classes of transfer operators: SM sphalerons, exotic sphalerons of 
a new gauge group and lepton or baryon number violation higher order operators. Since the DM is heavy it will be difficult to search for it but new particles at the weak or TeV scale are can be probed in collider experiments.

\section{Non-thermal Dark Matter Models}
\label{sec:ntdm}

\subsection{Cladogenesis} 
\label{cladogenesis}

Cladogenesis  \cite{Allahverdi:2010rh} is based on the observation that the dilution factor due to entropy release by moduli decay is very close to the observed baryon asymmetry. Indeed for a modulus $\tau$ with a reheating temperature in the range MeV -- GeV (corresponding to $M_\tau$ of order $20 - 1000$\,TeV) the dilution factor is given by 
\begin{equation}
Y_\tau=\frac{ 3 T_{RH} }{ 4 M_\tau^4 }\sim 10^{-9}-10^{-7}
\end{equation}
a value that is close to $\eta_B$ and also to $Y_{DM}$ as long as $M_{DM}$ is within a factor or two from the proton mass. At the same time any previous DM abundance will be suppressed by the same factor. These considerations lead the authors of Cladogenesis to consider a non-thermal origin of DM from modulus decay. The scenario goes as follow: $\tau$ decays to some species $N$ (\textbf{ 1$_I$}) and to DM (directly or via dark sector particles following \textbf{ 1$_D$}). The decay to DM must be suppressed down to $10^{-3}$ to achieve the observed relic abundance. $N$ then decays to SM by violating baryon (or lepton) number and CP to produce the correct baryon asymmetry (\textbf{ 2$_V^*$}). Note that the DM is not asymmetric in this model because baryogenesis is done in the visible sector only. 

Another example of non-thermal mechanism is given in \cite{Kohri:2009yn} where the  DM arises from the out-off equilibrium decay of the inflaton instead of the moduli.

\begin{table}[h]
\begin{center}
\begin{tabular}{|l|c|c|c|c|c|c|}
\hline
Model & DM& HS & BAU & $\mathcal{O}(M_{DM})$& Signal & Diagram \\
\hline
Two singlets EWBG\cite{Ahriche:2013zwa} &$\mathcal{WIMP}$  & \no&  EWPHT& $2 - 225$ GeV &DD-ID-CO& $5^*$\\
\hline
EW cogenesis \cite{Cheung:2013dca} & $\mathcal{WIMP}$ & \no&  EWPHT& GeV-TeV & CO& $5^*$\\
\hline
WIMPy L$^{(\dagger)}$\cite{Cui:2011ab}&$\mathcal{WIMP}$ & \no &  ANNIH &TeV &ID-CO & $T$-$4^*$  \\
\hline
WIMPy Q$^{(\dagger)}$\cite{Cui:2011ab}&$\mathcal{WIMP}$ & \no &  ANNIH &500GeV &DD-ID-CO & $T$-$4^*$  \\
\hline
Meta-stable WIMP\cite{Cui:2012jh}&$\mathcal{WIMP}$ & \no &  DECAY &GeV-TeV &CO & $T$-$4^*$  \\
\hline
Kitano-Low  \cite{Kitano:2004sv}&$\mathcal{ADM}$  & \yesno &DECAY&GeV & CO & $1_V^*$-$1_I^*$-$T$-$2_D$\\
\hline
Hylogenesis  \cite{Davoudiasl:2010am}&$\mathcal{ADM}$ & \yes & DECAY& 5 GeV& IND-DD & ${1}_V^*$-$1_D^*$-$T $ \\
\hline
ADM Leptog \cite{Falkowski:2011xh}& $\mathcal{ADM}$ & \yes  &  DECAY &KeV--10\,TeV& DD-ID & $1_V^*$-$1_D^*$-$ T$ \\
\hline
Darkogenesis \cite{Buckley:2010ui} &$\mathcal{ADM}$ & \yes     & TRANS&$5 - 15$ GeV& GW & $*$-$3$-$*$\\
\hline
Baryogenesis from DM \cite{Feng:2013wn} &$\mathcal{ADM}$ & \yes     & TRANS&$3$ GeV& DD-CO & $1_D*$-$3$-$*$\\
\hline
Aidnogenesis \cite{Blennow:2010qp} &$\mathcal{ADM}$ & \yes &DECAY&6 GeV &DD-FCNC -CO   & ${1}_V^*$-$\overline{3}$-$*$-$T$\\
\hline
Xogenesis \cite{Buckley:2010ui} &$\mathcal{ADM}$ & \yes   & TRANS&100GeV--TeV&CO &$*$-$3$-$*$ \\
\hline
Pangenesis \cite{Bell:2011tn} &$\mathcal{ADM}$ & \yes   & AFDIN&1.6--5 GeV& DD-CO &$*$-$1_I^*$-$T$-$2_V$-$2_D$\\
\hline
Cladogenesis \cite{Allahverdi:2010rh} &$\mathcal{NTDM}$  & \yesno  &  DECAY&  5-500GeV& - &$1_I$-$1_D$-${2}_V^* $\\
\hline
\end{tabular}
\end{center}
\caption{Summary of the models presented in this review and others. The first column shows the type of the DM candidate : WIMP, ADM or NTDM (Non--thermal DM).
The second one is about the hidden sector (HS): \yes \, means HS is necessary, \no \, means the model does not rely on HS and \yesno \, means that there are realizations of the idea with HS. The third column shows the mechanism that produces the observed baryon asymmetry: EWPHT (Electroweak Phase Transition), DECAY, ANNIH (Annihilation), TRANS (Transfer) or AFDIN (Affleck--Dine). The fourth column gives the order of magnitude of (the sum of) the mass of DM candidate(s). The column Signal shows the  predictions of the models: DD (Nuclear recoil direct detection experiments), ID (photons and/or neutrinos indirect detection experiments), CO (Collider), GW (Gravitational waves), IND (Induced proton decay) and FCNC (flavor changing neutral current). Finally the Diagram column shows the diagrammatic route followed by the model following the notation of Fig. \ref{fig:diag} (see text for details). 
($\dagger$) $L$ refers to the WIMPy baryogenesis scenario applied to leptons, whereas $Q$ refers to the baryonic version.}
\label{table}
\end{table}

%
\section{Summary}

In this short review we have given an overview of the models linking the generation of the baryon asymmetry of the universe and dark matter. These models are varied and diverse and tackle the problematic from different points of view. Models attempting to preserve the WIMP miracle lead to a very rich phenomenology and their couplings can be probed at LHC soon. These models do not address the coincidence between the baryon and DM asymmetries and the link between the two abundances is not strong. ADM models, one the other hand, give a natural explanation to this ratio at the price of WIMP phenomenology. Lastly non-thermal production models are yet another possibility relating the genesis of the dark and visible sector. 
LHC and dark matter search experiments will probe large chunks from the theoretical landscape of DM, hopefully shedding light on its nature and on the mechanism at work for baryogenesis.

We summarize the models discussed here in Table \ref{table} where we give information about the nature of their DM, the BAU mechanism at work, the existence of a hidden sector, range of the DM mass allowed in the model as well as the expected signal. The last column shows the diagrammatic signature of the model based on Fig.\ref{fig:diag} and the convention outlined in the introduction. 

\vskip15.mm

\section*{Acknowledgments}
S.M. thanks to DFG grant WI 2639/4-1 for financial support. S.B. was supported by the Spanish MINECO under grants FPA2011-22975 and MULTIDARK CSD2009-00064 (Consolider-Ingenio 2010 Programme), by Prometeo/2009/091 (Generalitat Valenciana), by the EU ITNUNILHC PITN-GA-2009-237920.

\bibliographystyle{unsrt}
\bibliography{minirev}
\end{document}